\documentclass[twocolumn]{jetpl}

\usepackage{epsfig}
\usepackage{amssymb}
\usepackage{amsmath}
\usepackage{amsfonts}
\usepackage{epstopdf}
\usepackage{bm}
\usepackage{cite}

\begin{document}

\title{Temperature derivative of the chemical potential and its magnetooscillations in two-dimensional system}

\rtitle{Temperature derivative of the chemical potential and its magnetooscillations in two-dimensional system}

\sodtitle{Temperature derivative of the chemical potential and its magnetooscillations in two-dimensional system}

\author{Y.~Tupikov$^{\triangle,+}$,A.Yu.~Kuntsevich$^{+,*}$\thanks{e-mail: alexkun@lebedev.ru},
V.M.~Pudalov$^{+,*}$, I.S.~Burmistrov$^{\ddag,*}$}
\rauthor{Y.~Tupikov, A.Yu.~Kuntsevich,
V.M.~Pudalov, I.S.~Burmistrov}
\sodauthor{Y.~Tupikov, A.Yu.~Kuntsevich,
V.M.~Pudalov, I.S.~Burmistrov}

\address{$^{+}$ P.N. Lebedev Physical Institute of the RAS, 119991 Moscow, Russia \\
          $^{*}$ Moscow Institute of Physics and Technology, 141700 Moscow, Russia\\
          $^{\triangle}$ Department of Physics, Pennsylvania State University, University Park, PA 16802, USA\\
          $^{\ddag}$ L.D. Landau Institute for Theoretical Physics, Kosygina
  street 2, 119334 Moscow, Russia
          }

\abstract{
 We report first thermodynamic measurements of the  temperature derivative of chemical potential ($\partial \mu/\partial T$) in two-dimensional (2D) electron systems. In order to test the  technique we have chosen Schottky gated GaAs/AlGaAs heterojunctions and detected experimentally in this 2D system quantum magnetooscillations of $\partial \mu/\partial T$. We also  present a Lifshits-Kosevitch type theory for the $\partial \mu/\partial T$ magnetooscillations in 2D systems and compare the theory with experimental data. The magnetic field dependence of the $\partial \mu/\partial T$ value appears to be sensitive to the density of states shape of Landau levels.  The data in low magnetic field domain demonstrate brilliant agreement with theory for non-interacting Fermi gas with Lorentzian Landau level shape.
 }


\dates{\today}{*}

\maketitle

 Quantum oscillations are known to be a universal tool to study the electron energy spectrum (Fermi surface cross-sections, electron effective mass and g-factor)  in three-dimensional single crystals and two-dimensional (2D) systems. In  contrast to the 3D-case, the 2D systems allow {\it in-situ} tuning the spectrum and the Fermi energy by various methods (including  electric field effect in gated structures, illumination, uniaxial stress etc.),  and, hence, allow
comprehensive magnetooscillation studies.
Quantum oscillations in 2D systems are most often studied in resistivity (Shubnikov-de Haas effect) \cite{ando}, and magnetization (de Haas-Van Alphen) \cite{Usher}. However, other physical quantities in 2D systems, such as thermo-EMF \cite{Fletcher}, heat capacity \cite{Wang},  chemical potential \cite{pudalovChempot, pudalovE_F}, compressibility \cite{eisenstein}, also oscillate with magnetic field.
All experimental methods applied for  measuring magnetooscillations of various observables have limitations for a practical use, therefore developing new magnetooscillations tools is of substantial importance.

All methods for chemical potential measuring  are based on the W.~Thomson (Kelvin) idea \cite{Kelvin} that if the two plates of a capacitor are made of different materials (with different work functions $\mu_{1,2}$), the charge of the plates is $C(\mu_1-\mu_2)/e$, where $C$ is the electric capacitance. Correspondingly, when the two plates are connected electrically and  an external parameter varies affecting one of the chemical potential values, a recharging current starts flowing between the plates. The recharging current is proportional to $\Delta\mu$ (in case the capacitance $C$ varies), $\propto \partial \mu/\partial n$ (in case one of the plates is a 2D gas of a density $n$   varying with  a gate voltage \cite{pudalovE_F, eisenstein}), $\propto \partial \mu/\partial B$ (in case the magnetic field $B$ varies\cite{teneh}), or $\propto \partial \mu/\partial T$ (in case the temperature varies \cite{nizhan}).

In our study we apply the technique of measuring $\partial \mu/\partial T$ similar to that used by Nizhankovskii for bulk samples \cite{nizhan}, to the 2D electron systems in magnetic field. We focus on $\partial \mu/\partial T$ oscillations
to compare them with semiclassical theory, and to determine the shape of the density of states of Landau levels. The advantage of the temperature modulation technique is the absence of eddy currents or a background signal (concomitant of many other techniques), and its pure thermodynamic origin.

\begin{figure}
\vspace{0.1 in}
\centerline{\psfig{figure=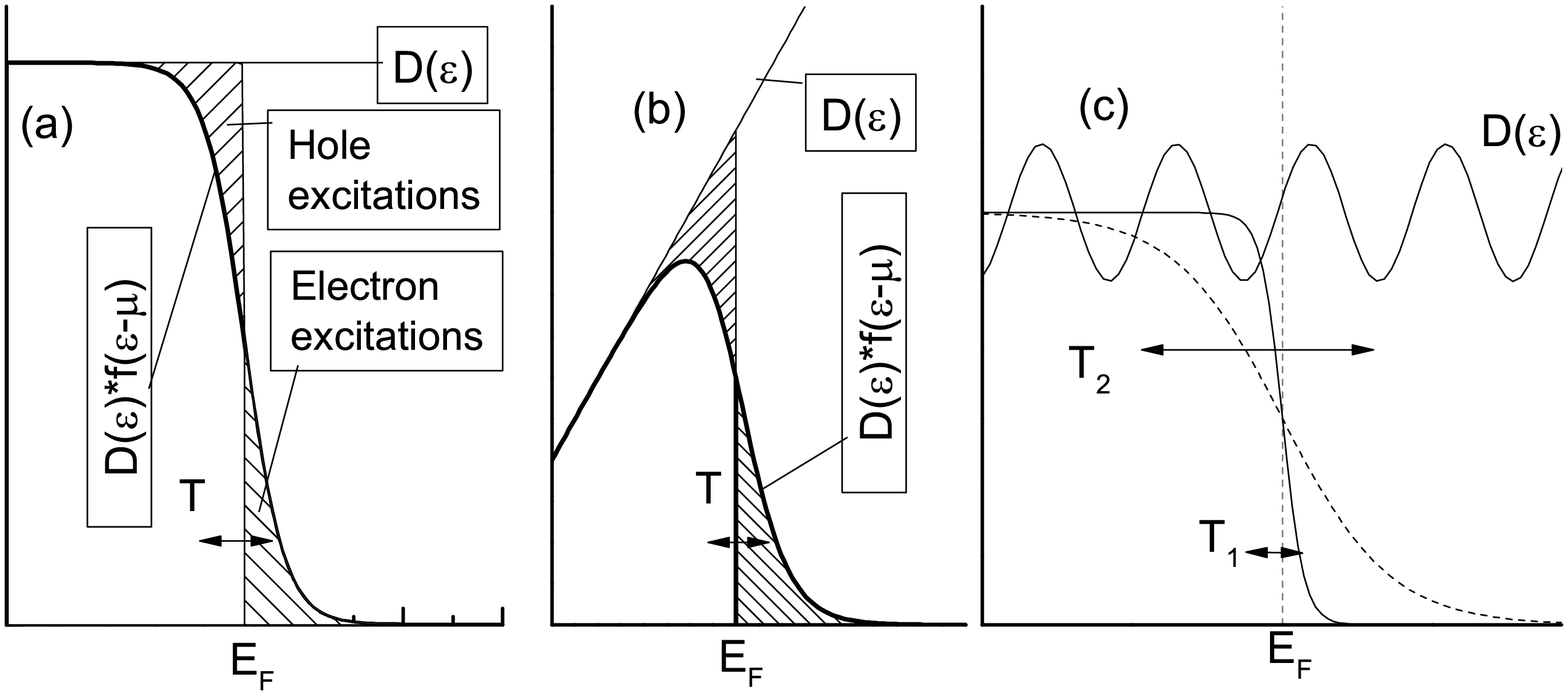,width=230pt}}
\begin{minipage}{3.2in}
\caption{(a) Density of states $D(\varepsilon)= const$ and its product with the Fermi distribution function. Equivalent hatched areas denote electron an hole excitations. (b) The same as (a) for the density of states changing with energy; the hatched areas are unequal. (c) Oscillatory density of states in perpendicular magnetic field and Fermi function corresponding to low temperature ($T<\hbar\omega_c,\Gamma$, solid line) and high temperature ($T<\hbar\omega_c,\Gamma$, dashed line).}
\label{fig0}
\end{minipage}
\end{figure}

{\it Qualitative discussion. -- \,} It is worthwhile to give a qualitative explanation why $\partial \mu/\partial T$ oscillates with perpendicular magnetic field. For the bare quadratic energy spectrum, $\varepsilon(p)=p^2/2m$, the single electron density of states is constant in two dimensions. At temperatures $T \ll E_F$ the number of particle-like excitations above $\mu$ equals to the number of hole-like excitations below $\mu$ (hatched areas in Fig. \ref{fig0}a). Therefore, for a fixed electron density $n$ the chemical potential is independent of temperature, $\partial \mu/\partial T=0$, with exponential accuracy at low temperatures, $T \ll E_F$, where $E_F$ stands for the Fermi energy.  In the case of energy dependent density of states ({like e.g. in 3D systems, graphene or 2D systems in quantizing magnetic field}), one can expand it in the vicinity of the Fermi energy (see Fig.\ref{fig0}b): $D(\varepsilon)=D(E_F)+ (\partial D(\varepsilon=E_F)/\partial \varepsilon)\times (\varepsilon-E_F)$
Then, {for a nonzrero  temperature}
the particle-hole asymmetry emerges (since the hatched areas do not coincide in Fig.\ref{fig0}b) and one needs to shift chemical potential in order to conserve the total number of particles. Using the standard low temperature expansion for the Fermi-type integrals (see, e.g., Ref.\cite{LL5}) one can easily find  correction to the chemical potential at low temperatures, $T\ll E_F$: $\mu(T)-E_F=- [\pi^2 T^2/6 D(E_F)]\times\partial D(\varepsilon=E_F)/\partial \varepsilon$. Hence, we find
\begin{equation}
 \left (\frac{\partial \mu}{\partial T}\right )_n =-\frac{\pi^2T}{3D(E_F)} \frac{\partial D(E_F)}{\partial E_F}=-\frac{\pi^2T}{3} \frac{\partial D(E_F)}{\partial n} .
\label{qualit}
\end{equation}
{ We note that this equation is applicable for degenerate Fermi-systems of any dimensionality and for any spectrum.}

{ In perpendicular field, due to Landau quantization $D(\varepsilon)$ becomes dependent on energy} 
(see Fig. \ref{fig0}c), and Eq.~(\ref{qualit}) should be averaged over $T$ in the vicinity of the Fermi energy. If the temperature is low ($T_1$, Fig.~\ref{fig0}c), one can directly apply Eq.~(\ref{qualit}); in the opposite limit of high temperature($T_2$, Fig.~\ref{fig0}c) the oscillations are averaged over a wide energy interval and  become exponentially suppressed.

Two qualitative predictions follow from the above considerations.
(i) In the maxima and minima of the density of states the $\partial \mu/\partial T$ signal is zero, and vice versa,  the signal is maximal where the derivative $\partial \ln D(E_F)/\partial E_F$ is maximal. (ii) The amplitude of the $\partial \mu/\partial T$ magnetooscillations
is a non-monotonic function of temperature: at low temperatures $T\ll \omega_c,\Gamma$ the $\partial \mu/\partial T \propto T$, whereas for high temperatures $T\gtrsim \omega_c$ averaging over several oscillations suppresses the signal.

{\it Theoretical background. -- \,} The thermodynamic potential of non-interacting electron system can be written as \cite{LL5}
\begin{equation}
\Omega(T,\mu,B) = -T \int d\varepsilon\, D(\varepsilon) \ln \left [ 1+e^{(\mu -\varepsilon)/T} \right ] .
\label{eq:Omega:B}
\end{equation}
In the presence of the perpendicular magnetic field $B$
the density of states becomes
\begin{equation}
D(\varepsilon) =  \frac{m\omega_c}{2\pi}\sum_{\sigma=\pm} \sum_{n=0}^\infty  \mathcal{W}\bigl (\varepsilon+\sigma Z -\omega_c(n+1/2)\bigr ) ,
\end{equation}
where $Z=g_L \mu_B B/2$ ($g_L$ stands for the $g$-factor) describes the effect of the Zeeman splitting. The function $\mathcal{W}(\varepsilon)$ describes broadening of a Landau level due to disorder. It satisfies the normalization condition:
$\int d\varepsilon\, \mathcal{W}(\varepsilon) = 1$. We note that, in general, this function can be different for different Landau levels. Using the function $\mathcal{W}(\varepsilon)$ a typical width of a Landau level can be estimated as $\Gamma \sim \left [  \int d\varepsilon\,  \mathcal{W}^{\prime\prime}(\varepsilon) \right ]^{-1/2}$. The accurate quantum-mechanical evaluation of $\mathcal{W}(\varepsilon)$ for a given type of disorder is a complicated problem and solved only partially \cite{wegner,big,il,affleck,em,bs,gd}. In the absence of disorder, one obviously has $\mathcal{W}(\varepsilon) = \delta(\varepsilon)$. There are three commonly used models of the disorder Landau level broadening \cite{ando,KMT}:
\begin{align}
\textrm{Lorentzian model: }\quad
\mathcal{W}(\varepsilon) & = \frac{1}{\pi} \frac{\Gamma}{\varepsilon^2+\Gamma^2} \notag
\\
\textrm{Gaussian model:}\quad
\mathcal{W}(\varepsilon) & = \frac{1}{\Gamma\sqrt{\pi}} e^{-\varepsilon^2/\Gamma^2}
\label{eq:def:W}\\
\textrm{semicircle model:}\quad
\mathcal{W}(\varepsilon) & = \frac{2}{\pi \Gamma} \sqrt{1-\varepsilon^2/\Gamma^2}
\notag
\end{align}

In a standard way, by means of the Poisson resummation formula applied to the thermodynamic potential  \eqref{eq:Omega:B} we obtain the Lifshitz-Kosevich-type \cite{LK} expression for $\partial \mu/\partial T$:
\begin{align}
\left ( \frac{\partial \mu}{\partial T} \right )_n & = -\sum_{k=1}^{\infty}
  \frac{2 \pi  (-1)^{k} \mathcal{A}_k}{\sinh \mathcal{X}_k}\Bigl [ 1 -\mathcal{X}_k \coth \mathcal{X}_k \Bigr ]    \sin \frac{2\pi \mu k}{\omega_c} \notag \\
& \times \cos \frac{2\pi Z k}{\omega_c} \Biggl [ 1 +  2 \sum_{k=1}^{\infty}
 \frac{(-1)^{k} \mathcal{A}_k \mathcal{X}_k}{\sinh \mathcal{X}_k} \cos \frac{2\pi \mu k}{\omega_c}\notag \\
& \hspace{2cm}\times \cos \frac{2\pi Z k}{\omega_c}\Biggr ]^{-1}
 \, .
\label{form1}
\end{align}
This result holds for $\mu\pm Z \gg \omega_c, T, \Gamma$.
Here $\mathcal{X}_k= 2\pi^2 T k/\omega_c$, and
\begin{equation}
\mathcal{A}_k  = \int d\varepsilon\, \mathcal{W}(\varepsilon) \exp\left ( \frac{2\pi i\varepsilon k}{\omega_c} \right )\,
\end{equation}
characterizes damping of oscillations due to Landau level broadening. For three models in Eq. \eqref{eq:def:W} it becomes
\begin{align}
\textrm{Lorentzian model: }\quad
\mathcal{A}_k & = e^{-2\pi \Gamma |k|/\omega_c} \notag
\\
\textrm{Gaussian model:}\quad
\mathcal{A}_k & = e^{-\pi^2 \Gamma^2 k^2/\omega^2_c}
\label{eq:def:A}\\
\textrm{semicircle model:}\quad
\mathcal{A}_k & = \frac{\omega_c}{\pi \Gamma k} J_1\left (\frac{2\pi \Gamma k}{\omega_c}\right )
\notag
\end{align}
where $J_1(x)$ denotes the Bessel function.

For low temperatures, $T\ll \omega_c, \Gamma, \mu\pm Z$, the $\partial \mu/\partial T$ value is given by the result \eqref{qualit} with $D=(1/2)\sum_{\sigma=\pm}D(E_F^\sigma)$. Here $E_F^\sigma = 2\pi n_\sigma/m$ denotes the Fermi energy for a a given spin projection. The corresponding electron density is determined by the Zeeman splitting, $n_\sigma/n=1+\sigma Z/E_F$ where $E_F=2\pi n/m$. In the case of the Lorentzian broadening of a Landau level the density of states at the Fermi energy becomes
\begin{equation}
D(E_F^\sigma) = \frac{m}{\pi} \frac{(1/2) \sinh (2\pi \Gamma/\omega_c)}{\sinh^2(\pi\Gamma/\omega_c)+\cos^2(\pi E_F^\sigma/\omega_c)}
 \,  .
 \label{eq:S:T0:H}
\end{equation}

We mention that above we ignore the effect of electron-electron interaction. As well-known,
for an interacting 3D electron system the Lifshitz-Kosevich-type expressions for magnetooscillations of the thermodynamic potential can be obtained from the non-interacting expressions via standard Fermi-liquid renormalization of  the quasiparticle spectrum \cite{L, BG}. In 2D, it is not the case, in general \cite{CS}. For classically weak perpendicular magnetic field, modification of the Lifshitz-Kosevich-type expression for a 2D electron interacting disordered system has been studied in Refs. \cite{MMR,AGM}.

{\it Experimental results. -- \,} We studied two gated GaAs/AlGaAs single heterojunctions (GaAs1 and GaAs2, similar to tat from Ref\cite{mokherov}) in the temperature range between 2.5 and 25K with electron mobilities 20 and 25 $m^2/Vs$, respectively.  The modulation technique was similar to that used by Nizhankovskii\cite{nizhan} and will be reported elsewhere\cite{tupikov}.
The density could be varied in the interval from $2\cdot 10^{11}$cm$^{-2}$ to $4\cdot 10^{11}$cm$^{-2}$ by changing the gate voltage. The samples had  5\,mm$^2$ area, the gate-to-2D gas capacitance was 1100 pF. Both samples demonstrated similar results, we present therefore the data for GaAs2 only.

\begin{figure}
\vspace{0.1 in}
\centerline{\psfig{figure=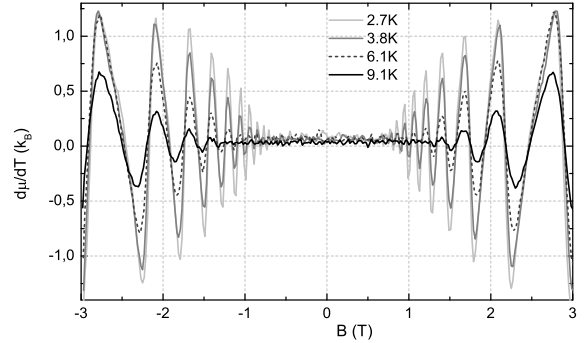,width=230pt}}
\begin{minipage}{3.2in}
\caption{Fig.~\ref{fig1}. $\partial \mu/\partial T$ versus magnetic field in sample GaAs-2 at various temperatures. $n=4.16 \cdot 10^{11}$cm$^{-2}$ }
\label{fig1}
\end{minipage}
\end{figure}

\begin{figure}
\vspace{0.1 in}
\centerline{\psfig{figure=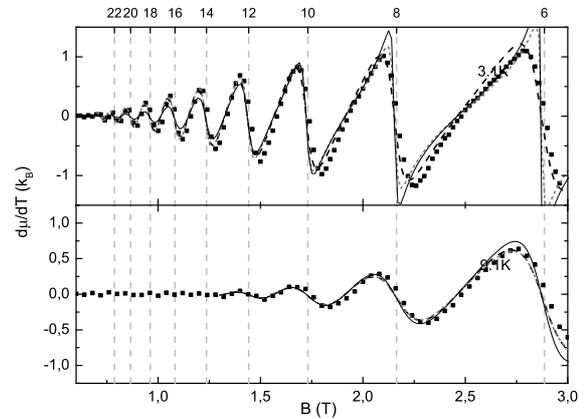,width=230pt}}
\begin{minipage}{3.2in}
\caption{Fig.~\ref{fig2}. $\partial \mu/\partial T$ versus magnetic field in sample GaAs-2 at $n=4.16 \cdot 10^{11}$cm$^{-2}$. Data is fitted with Lorentzian (dashed curves),Gaussian (solid curves) and Gaussian with $\Gamma\propto\sqrt{B}$ (dotted curves) lineshapes at 3K (panel a) and 9K (panel b). Upper axis shows filling factor. }
\label{fig2}
\end{minipage}
\end{figure}

Examples of low-field $\partial \mu/\partial T$ oscillations measured for the sample GaAs-2 are shown in Fig.~\ref{fig1}. For fields lower than 3 Tesla and in the range of temperatures 2.7-9.1K we fitted  the data  using Eq.~\eqref{form1}. We use the bare band mass ($m_b=0.067m_e$ for GaAs).  Because of the low-field range studied and large cyclotron splitting, all the results below are insensitive to the g-factor value  (that may vary in the range $g=-0.4\div2$).  The best fit is obtained for the broadening of Landau levels described by the Lorentzian model with the width $\Gamma$=0.4\,meV which is independent of $B$ (see dashed curves in Fig.~\ref{fig2}). The range of temperatures ($0.22$mV$<T<$0.8 mV) and magnetic fields ($0.86<\hbar\omega_c<5.2$ mV) used for fitting procedure is wide enough to identify  the line shape with the Loretzian curve.  On the other hand,  the Gaussian model  with $\Gamma$  independent of $B$ (solid curves in Fig.~ \ref{fig2}) fails to fit the oscillations. The Gaussian model with level broadening proportional to $\sqrt{B}$ works only in low magnetic fields (dotted curves in Fig. \ref{fig2}). We checked that in the density range from 3$\cdot 10^{11}$\,cm$^{-2}$ to 5$\cdot 10^{11}$\.cm$^{-2}$ the low-field level broadening is constant (equal to 0.4 meV for sample GaAs-2). The Lorentzian lineshape of the Landau levels is in the agreement with magnetization measurements of Potts et al.~\cite{Potts} on the moderate mobility GaAs-based sample.
Generally, the shape of DOS  in 2D systems is still debated, with a history of research reviewed, e.g., by Usher et al.~\cite{Usher}. Different approaches are used to determine DOS with many contradictory results obtained.

{\it Conclusions. -- \,}
{ We presented here experimental test  of the novel technique of thermodynamic $\partial \mu/\partial T$ measurements, that is ideally suitable for 2D gated carrier systems. The technique  appears to be highly sensitive and thereby enables thermodynamic measurements with a single-layer electron system comprising only $10^{10}$ electrons. We also  present a Lifshits-Kosevitch type calculations for the $\partial \mu/\partial T$ magnetooscillations in 2D systems and compare the theory with experimental data; the comparison reveals a good agreement between the data and the theory. The magnetic field dependence of the $\partial \mu/\partial T$  appears to be rather sensitive to the shape of the density of states  at Landau levels, and in the particular case of moderate mobility samples  studied we found the density of states to correspond to the Lorentzian rather than Gaussian curve.
We want to emphasize, that the method described here suggests an independent opportunity of extracting energy spectrum of the 2D carrier system.}

\vspace{1cm}
{\bf Acknowledgments}\\
AYuK acknowledges support by Russian Ministry of science and education (President Grant MK-4208.2013.2),  VMP and ISB acknowledge support by RSCF (14-12-000879).

\end{document}